\documentclass[conference, a4paper]{IEEEtran}
\IEEEoverridecommandlockouts
\usepackage{mathtools}
\usepackage{amssymb,amsmath}
\usepackage{ifpdf}
\usepackage{url}
\usepackage{graphicx}
\usepackage{caption}
\usepackage{subcaption}
\usepackage{cite}
\usepackage{etoolbox}
\usepackage{multirow}
\usepackage{tabu}
\usepackage{wrapfig}

\usepackage{array}
\newcolumntype{L}[1]{>{\raggedright\let\newline\\\arraybackslash\hspace{0pt}}m{#1}}
\newcolumntype{C}[1]{>{\centering\let\newline\\\arraybackslash\hspace{0pt}}m{#1}}
\newcolumntype{R}[1]{>{\raggedleft\let\newline\\\arraybackslash\hspace{0pt}}m{#1}}

\usepackage{algpseudocode}

\algnewcommand\algorithmicswitch{\textbf{switch}}
\algnewcommand\algorithmiccase{\textbf{case}}
\algnewcommand\algorithmicassert{\texttt{assert}}
\algnewcommand\Assert[1]{\State \algorithmicassert(#1)}%
\algdef{SE}[SWITCH]{Switch}{EndSwitch}[1]{\algorithmicswitch\ #1\ \algorithmicdo}{\algorithmicend\ \algorithmicswitch}%
\algdef{SE}[CASE]{Case}{EndCase}[1]{\algorithmiccase\ #1}{\algorithmicend\ \algorithmiccase}%
\algtext*{EndSwitch}%
\algtext*{EndCase}%

\usepackage[capitalise]{cleveref}

\RequirePackage{pgfplots}
\pgfplotsset{compat=newest}
\usetikzlibrary{plotmarks}
\usetikzlibrary{arrows.meta}
\usetikzlibrary{shapes}
\usepgfplotslibrary{units}
\usepgfplotslibrary{patchplots}
\usepackage{tikzscale}
\usepackage{tikz}
\usetikzlibrary{arrows,positioning,shapes.geometric,arrows.meta,decorations.markings} 

\tikzstyle{basic box}[black] = [shape=rectangle, align=center, draw=#1, rounded corners] 
\tikzstyle{basic box b}[red] = [shape=rectangle, align=center, draw=#1, rounded corners]
\tikzstyle{state} = [circle, draw, very thick]

\tikzstyle{basic box b}[red] = [shape=rectangle, align=center, draw=#1, rounded corners]

\tikzset{
    >=stealth',
    punkt/.style={
           rectangle,
           rounded corners,
           draw=black, very thick,
           text width=6.5em,
           minimum height=2em,
           text centered},
    pil/.style={
           ->,
           thick,
           shorten <=2pt,
           shorten >=2pt,},
    flash/.style args={#1:#2}{postaction=decorate,decoration={name=markings,
    mark=at position #1 with {%
    \draw[fill=#2, line width=.75\pgflinewidth, line cap=round, line join=round]
         (+\pgflinewidth,+7\pgflinewidth)   -- ++ ( left:+2\pgflinewidth) 
      -- (+-4\pgflinewidth,+-\pgflinewidth) -- ++ (right:+5\pgflinewidth)
      -- (+-\pgflinewidth,+-7\pgflinewidth) -- ++ (right:+2\pgflinewidth)
      -- (+4\pgflinewidth,\pgflinewidth)    -- ++ (left:+5\pgflinewidth)
      -- cycle;}}}
}

\begin{document}
%
\title{Modeling and Optimization of a Longitudinally-Distributed Global Solar Grid}


%
\author{\IEEEauthorblockN{Harsh Vardhan, Neal M Sarkar, Himanshu Neema}
\IEEEauthorblockA{Institute for Software Integrated Systems, Vanderbilt University, Nashville, TN, USA
}}


\maketitle


\subsection*{\textbf{\textit{ABSTRACT}}}
\label{sec:abstract}
\textbf{
Our simulation-based experiments are aimed to demonstrate a use case on the feasibility of fulfillment of global energy demand by primarily relying on solar energy through the integration of a longitudinally-distributed grid. These experiments demonstrate the availability of simulation technologies, good approximation models of grid components, and data for simulation. We also experimented with integrating different tools to create realistic simulations as we are currently developing a detailed tool-chain for experimentation. These experiments consist of a network of model houses at different locations in the world, each producing and consuming only solar energy. The model includes houses, various appliances, appliance usage schedules, regional weather information, floor area, HVAC systems, population, number of houses in the region, and other parameters to imitate a real-world scenario. Data gathered from the power system simulation is used to develop optimization models to find the optimal solar panel area required at the different locations to satisfy energy demands in different scenarios.
}
\subsection*{Keywords}
\textbf{\textit{ Solar Energy, Global Grid, Optimization, Model-Based Simulation, Data-Driven Simulation, Cyber-Physical System, GridLAB-D, Sustainable Future.}}
\section{Introduction}
\setlength{\parskip}{0pt}
\setlength{\parsep}{0pt}
\setlength{\headsep}{0pt}
\setlength{\topskip}{0pt}
\setlength{\topmargin}{0pt}
\setlength{\topsep}{0pt}
\setlength{\partopsep}{0pt}
\label{sec:introduction}
Solar power generation from photo-voltaic (PV) cells has been rapidly growing, with global PV capacity steadily increasing over past years. Ongoing decreases in the cost of solar panel production, as well as increases in solar panel efficiency, provide additional incentives for energy producers to shift toward PV technologies \cite{barbose2016tracking}.
It seems reasonable that with a growing global population, depleting non-renewable resources, and availability of solar energy ($10^4$ times of current global energy needs), future generations will rely heavily on solar energy\cite{kuwano1994pv}\cite{europe2016global}. 
While solar power systems offer a compelling renewable source of energy, they suffer from reliability issues stemming from the intermittent availability of solar resources. In particular, solar production is limited to daylight hours, and local meteorological and geographic factors affect production.

Users of such a system rely on energy storage technologies or existing grid infrastructure to meet part of their demand. Transitioning to solar energy requires a reliable solar system that continuously balances energy production with consumption. High voltage DC (HVDC) and Ultra-High voltage DC (UHVDC) transmission systems can enable transcontinental exchange of power between distributed participant regions\cite{rudervall2000high}. 
Extending PV systems to a global, longitudinally-distributed (LD) scale stabilizes solar energy availability and reduces the need for energy storage and non-renewable energy production. An optimized solar power grid system ensures that energy production closely matches consumption, thereby minimizing solar installation area, maintenance, and operational costs, while maintaining reliability. Modeling, simulation, and optimization of  longitudinally-distributed solar system can provide insights into the effects of weather, human consumption patterns, solar panel area requirements, grid integration models, and integrated systems' characteristics.

Global power models are trending toward grid integration. For example, energy providers in Europe and North Africa aim to integrate a super grid that can harvest wind energy from the North and Baltic Seas and solar energy from the North African region. In Asia, China, South Korea, and Japan are considering a super grid to encourage more renewable energy production. Such grid integration and other small networks provide the groundwork for implementing a global super grid\cite{rudenko1991possible}. However, many questions must first be answered, such as \textit{"Can renewable production adequately match energy demands?"}, \textit{"How should renewable resources be integrated?"},\textit{"How do user consumption patterns affect the grid?"} and \textit{"How can weather fluctuations be accounted for to ensure production stability?"}

This paper presents a tripartite experimental procedure for the modeling, simulation, and optimization of an LD solar system. The approach entails gathering the scattered solar energy distributed across locations (assuming availability of power transportation facilities), modeling realistic houses, solar panels, and usage schedules of appliances, as well as using weather data to derive consumption and production patterns. The rest of the paper is organized as follows: Section~\ref{sec:relatedWorks} provides an overview of related literature. Section~\ref{sec:methods} provides a description of the tools, models, and methods used for experiments. Section~\ref{sec:experimentalResults} presents simulation and optimization results. Section~\ref{sec:discussion} discusses the experiment's limitations, technical considerations, and other observations. Section~\ref{sec:conclusionFutureWork} concludes the paper and suggests directions for future work.
\section{Related Work}
\setlength{\parskip}{0pt}
\setlength{\parsep}{0pt}
\setlength{\headsep}{0pt}
\setlength{\topskip}{0pt}
\setlength{\topmargin}{0pt}
\setlength{\topsep}{0pt}
\setlength{\partopsep}{0pt}

\label{sec:relatedWorks}
\captionsetup{justification=raggedright,singlelinecheck=false}
An interconnected global energy grid, with power generation localized in remote solar and wind plants, has been suggested as a viable option to reduce storage requirements\cite{chatzivasileiadis2013global}. A control technique for switching between primary and secondary energy sources was described using a hybrid energy network which primarily relies on wind and solar sources, with a hydrogen fuel cell as an auxiliary source\cite{das2005optimal}. Optimization of the capacity of distributed energy resources in a renewable energy-based grid has been presented using a sizing model to determine optimal battery size for grid stabilization \cite{kaabeche2011sizing}. An international grid which satisfied base load power demands using distributed wind farm generation was described in\cite{archer2007supplying}, where the problem of intermittent availability of wind resources was considered, and an interconnected, distributed production scheme was offered as a viable solution to the grid stability problem. A very large-scale PV system, where production is centralized around a desert area with high solar irradiance, was evaluated for economic costs, and technologies were proposed that could increase production and distribution efficiency\cite{kurokawa2002cost}. Optimization of both solar and wind production using meteorological data was presented by developing a linear optimization model for a mostly renewable energy grid at both the European and global scales\cite{aboumahboub2010optimization}. Integration of standalone hybrid renewable resources and their optimization was also studied in \cite{tezer2017evaluation}.

\section{Modeling and Experiment Flow}
\setlength{\parskip}{0pt}
\setlength{\parsep}{0pt}
\setlength{\headsep}{0pt}
\setlength{\topskip}{0pt}
\setlength{\topmargin}{0pt}
\setlength{\topsep}{0pt}
\setlength{\partopsep}{0pt}

\setlength{\belowcaptionskip}{-10pt}
\captionsetup{justification=raggedright,singlelinecheck=false}

\begin{figure}[ht]
	\centering
	\captionsetup{justification=centering,labelsep=period}
	\includegraphics[width=.4\textwidth]{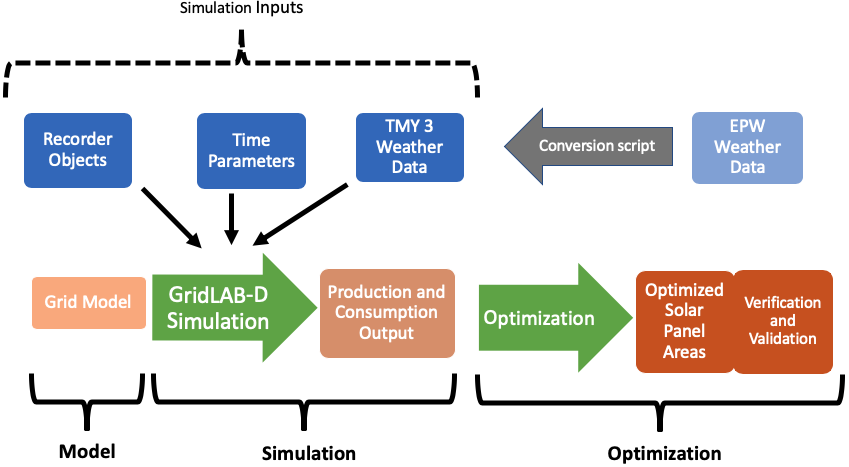}
	\caption{Experimental workflow}
	\label{fig:workflow}
\end{figure}

\label{sec:methods}
The overall experimental workflow, shown in Figure~\ref{fig:workflow}, consists of three major sections: modeling, simulation, and optimization. Other aspects of the workflow included data pre- and post-processing. The experiment was modeled in two stages. First, the grid equipment (e.g., house, solar panels, and appliances for each location) and distribution grid were modeled. The model was simulated using a power system simulator and energy consumed by households. Solar energy produced by panels based on weather data at each time-step was logged. Second, the optimization model was designed based on the model of grid integration. In the optimization step, data produced by the power system simulation was utilized. The experiment assumed the availability of a transmission grid and that its losses were negligible.

\subsection{ Longitudinally-Distributed Global Network} 
\begin{figure}[ht]
	\centering
	\captionsetup{justification=centering,labelsep=period}
	\includegraphics[width=.45\textwidth]{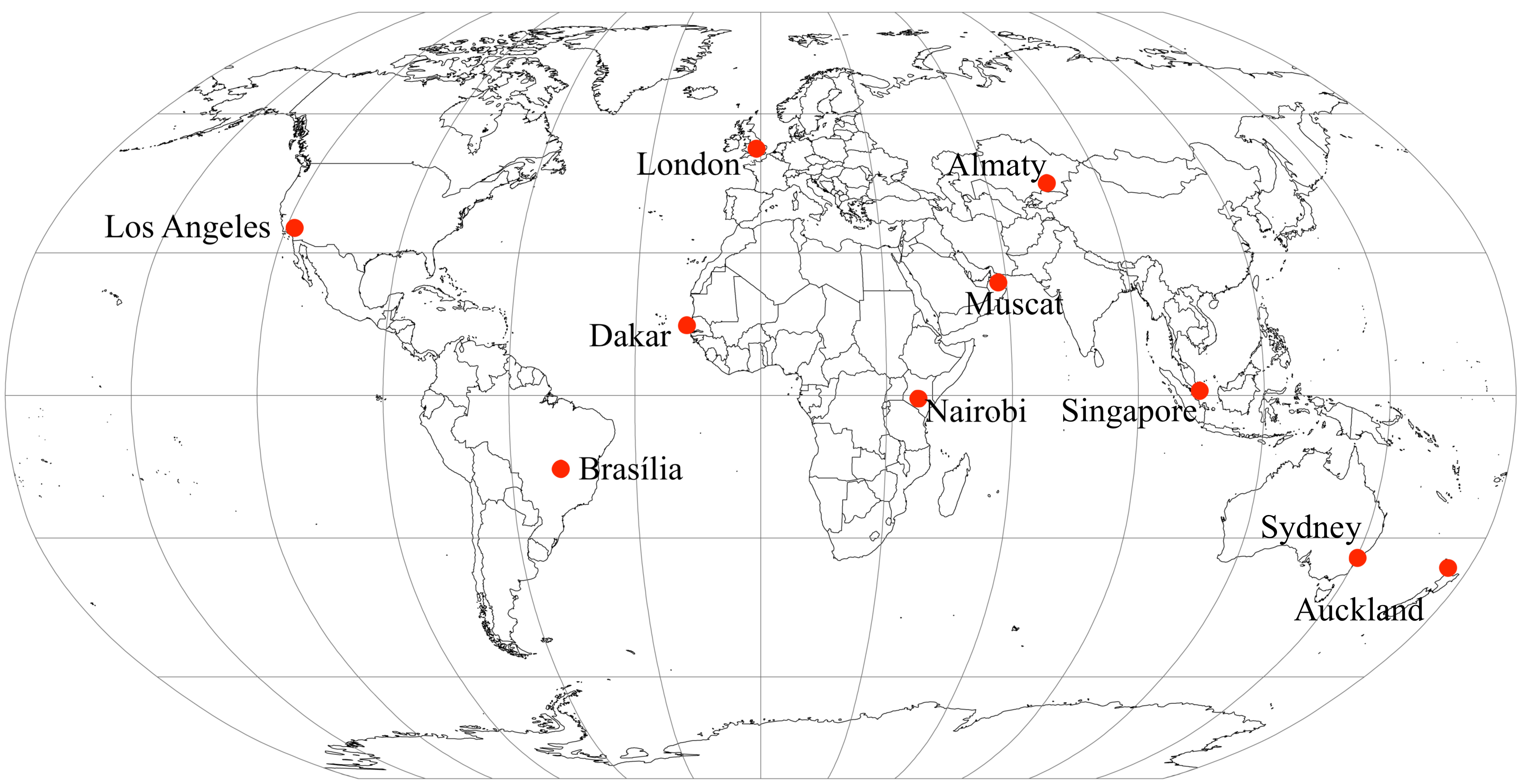}
	\caption{Map of experimental locations}
	\label{fig:map}
\end{figure}

Ten locations, shown in Figure~\ref{fig:map}, were selected for experimentation of a longitudinal global grid. Availability of grid connectivity at these locations was assumed. Selection of these locations was based on constraints that the sun should always shine at some location in network, thereby fulfilling all participants' energy needs. Selected locations were offset by approximately two time zones.

\subsection{Grid Model} 
Modeling of each location was generalized and represented by a model house with appliances. These houses were connected to form a small grid, but still used weather data from different locations and schedules. Each location also had a unitized solar panel attached to it. The model house had residential heating, ventilating, and air-conditioning (HVAC) systems, where temperatures could be specified as a schedule for any day. The HVAC system operated based on the thermostat schedule and weather of that location. Thermal performance of house was configured by the glazing effect, heat loss coefficient, interior mass surface conductance, and other factors. Heat gains from solar radiation and appliances were combined with those from the heating/cooling system to form the heat gains to the air. Houses also had a water heating system and a ZIP load. The ZIP load model, which consisted of  constant  impedance,  constant  current,  and  constant  power components,  represented the total appliance load. We defined the size of the water heater unit and ZIP-equivalent loads of all other appliances. Power consumed by the HVAC systems, water heaters, and ZIP loads depended on appliance schedules given as an input for $24$ hours. House power consumption also depended on the weather (e.g., air conditioners take more time and energy to cool the house in the summer than in the winter). Each location also had solar panels that produced power based on weather (e.g., solar radiation incident on solar panel and ambient temperature) and solar panel configurations (e.g., solar panel efficiency, material type, and tilt angle). We approximated these to real-world equipment parameters.

\begin{figure}[ht]
	\centering
	\captionsetup{justification=centering,labelsep=period}
	\includegraphics[width=.5\textwidth]{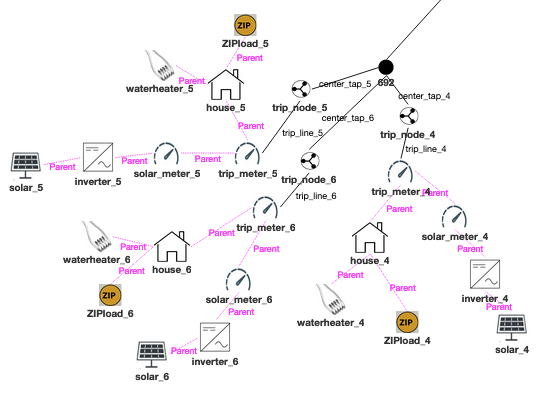}
	\caption{Partial grid model showing three residential modules}
	\label{fig:modelChunk}
\end{figure}

The grid was modeled using a distribution system modeling tool called the GridLAB-D Design studio, which is a web-based graphical modeling environment built using WebGME (Web-based Generic Modeling Environment)\cite{neema2019web}\cite{neema2019design}\cite{webgme}. WebGME is a web-based meta-modeling environment that enables creation of Domain-Specific Modeling Languages (DSMLs) and provides flexibility through its support for plugins that interpret the models and synthesize application artifacts, as well as through its programmable graphical user interface. The modeling environment utilizes Pacific Northwest National Laboratory's GridLAB-D simulator in the backend \cite{chassin2014gridlab}. A section of the model in GridLAB-D design studio environment is shown in Figure~\ref{fig:modelChunk}. 

\subsection{Weather Data and Schedules}
Weather data for each location were obtained from an open-source repository\cite{climate.onebuilding}, and then converted to the appropriate Typical Meteorological Year (TMY3) file format\cite{wilcox2008users} using a custom python script. A schedule consisted of values of associated variable parameters such as heating set point and water usage pattern, load on-off timing were created for each residential module to reflect real-life usage patterns.

\subsection{Model Transformation and Simulation}
The designed model was transformed by a model transformation tool in the GridLAB-D design studio, generating a script compatible with GridLAB-D\cite{neema2019web}. A recorder object logged (as .CSV files) the relevant production and consumption data during a 24-hour simulation with a time-step of 1 hour, from 01/02/2009, 0:00 hours to 01/02/2009, 24:00 hours. Power consumption between simulation steps was assumed to be constant.

\subsection{Integration and Optimization Model}
The output of the GridLAB-D simulation was the power consumption of the model houses at each location based on usage schedules and solar energy generated by the unitized panels. The consumption and production of each house was scaled by the number of houses at each location. For locations where information regarding quantity of houses was available, we used those data to scale consumption/production. For other locations, we used a house quantity equal to one-third of the population at those locations (a factor derived from empirically observed behavior at various locations).

The optimization problem was modeled and populated using the total consumption and unitized scaled solar production data. In experiment 1, we assumed that solar panels could be located at any of the ten locations. We had the constraint that the total area of panels across the grid should be distributed such that the total solar energy produced by solar panels would always meet the energy demand for all locations over the 24-hour simulation period. The goal of the optimization was to meet all regional energy needs while minimizing the total solar panel area required.

Given \textit{n} regions $R_1,\dotsc,R_n$ with consumption $C_{1,1},\dotsc,C_{24,n}$ and unitized solar productions $P_{1,1},\dotsc, P_{24,n}$. Let $A_1,\dotsc,A_n$ be the corresponding solar area scale factors. In our experiments, we used \textit{$n=10$}. The consumption and production is represented by $2-D$ matrix where the first subscript represents each hour of the simulation and the second subscript represents the index of a location. An algebraic representation of the optimization problem can be framed as follows:

\[
\begin{bmatrix}
P_{1,1}  & \dots & P_{1,n} \\
P_{2,1}  & \dotsc & P_{2,n}\\
\vdots  & \vdots & \vdots \\
P_{24,1}  & \dotsc & P_{24,n}\\
\end{bmatrix} *
\begin{bmatrix}
A_1\\
A_2\\
\vdots \\
A_n\\
\end{bmatrix} 
\geq
\begin{bmatrix}
\Sigma(C_{1,1}+ \dots +C_{1,n} )\\
\Sigma(C_{2,1}+ \dots +C_{2,n} )\\
 \vdots\\
\Sigma(C_{24,1}+ \dots +C_{24,n} )\\
\end{bmatrix}
\]

\[
\begin{bmatrix}
\Sigma(P_{1,1}+\dots +P_{24,1}) *A_1\\
\Sigma(P_{1,2}+\dotsc + P_{24,2})*A_2\\
\vdots   \\
\Sigma(P_{1,n}+\dotsc +P_{24,n})*A_n\\
\end{bmatrix}
\geq
0.4*
\begin{bmatrix}
\Sigma(C_{1,1}+\dots+C_{24,1} )\\
\Sigma(C_{1,2}+\dots+C_{24,2} )\\
 \vdots\\
\Sigma(C_{1,n}+\dots+C_{24,n} )\\
\end{bmatrix}
\]

\[
\begin{bmatrix}
\Sigma(P_{1,1}+\dots +P_{24,1}) *A_1\\
\Sigma(P_{1,2}+\dotsc + P_{24,2})*A_2\\
\vdots   \\
\Sigma(P_{1,n}+\dotsc +P_{24,n})*A_n\\
\end{bmatrix}
\leq
4*
\begin{bmatrix}
\Sigma(C_{1,1}+\dots+C_{24,1} )\\
\Sigma(C_{1,2}+\dots+C_{24,2} )\\
 \vdots\\
\Sigma(C_{1,n}+\dots+C_{24,n} )\\
\end{bmatrix}
\]
\[
\begin{bmatrix}
A_1\\
A_2\\
\vdots \\
A_n\\
\end{bmatrix} 
\geq
0
\]


This optimization model was created using the AMPL toolkit \cite{ample}. Scaled consumption and production data from each region at each time-step was provided to the solver, which then generated a solution. This solution was then validated by ensuring that the optimized production always exceeded aggregate consumption across the simulation (Figure ~\ref{fig:exp1}). 

In other three experiments, we configured each location to produce a minimum of 40\% (experiment 2), a maximum of 400\% (experiment 3), and 40-400\% (experiment 4) of its energy consumption, respectively. Additional constraints were added in AMPL for these cases (see Table~\ref{tab:optimization_results} for results).

For all experiments, the production of solar power was always higher than consumption by the households. This may be unwanted, so excess power produced by the solar panel can be either drawn from and used for another purpose, or a control mechanism can be deployed to control the active solar areas by aligning net production and consumption.
\section{Experimental Results}
\setlength{\parskip}{0pt}
\setlength{\parsep}{0pt}
\setlength{\headsep}{0pt}
\setlength{\topskip}{0pt}
\setlength{\topmargin}{0pt}
\setlength{\topsep}{0pt}
\setlength{\partopsep}{0pt}
\captionsetup{justification=raggedright,singlelinecheck=false}

\label{sec:experimentalResults}
The experiment outcome encompassed two results : \textbf{production and consumption data} from the power simulation and \textbf{panel area scale factors} from the optimization. These data from the two experiments are discussed below.

In experiment 1, the optimization problem was to find the minimum area at different locations to meet total energy consumption needs at all locations through solar energy for 24 hours, sampled at a one hour time step. There was no lower or upper limit on production capacity of each location, but the cumulative sum of production was required to be greater than cumulative consumption. In experiment 2, we removed this flexibility by adding another constraint on the amount of solar production at each location, which was required to be $40\%$ of its consumption. Optimization results of both experiments are shown in Table~\ref{tab:optimization_results}.

\begin{table}
\centering
\small
\begin{tabular}{ | L{1.5 cm} | L{1.2cm}| L{1 cm}| L{1.2 cm}|L{1.4 cm}| } 
\hline
\textbf{Location} & \textbf{No  constraint} & \textbf{ $>40\%$ } & \textbf{ $<400\%$ } & \textbf{$>40\%$ and $<400\%$}\\
\hline
Los Angeles & 36.78 & 34  & 66.39 & 60.08 \\
\hline
Bras\'{i}lia & 58.10 & 57.63 & 53.32 & 53.45 \\
\hline
Dakar & 0 & 4.41 & 0 & 4.41\\
\hline
London & 0 & 27.04 & 0 & 27.04\\
\hline
Nairobi & 23.85 & 5.08 & 26.38 & 7.31 \\
\hline
Muscat & 0 & 6.59 & 0 & 6.60\\
\hline
Almaty & 0 & 27.35 & 0 & 27.35 \\
\hline
Singapore & 4.19 & 11.99  & 15.06 & 21.57\\
\hline
Sydney & 66.90 & 64.27 & 44.82 & 44.82\\
\hline
Auckland & 0 & 5.44 & 0 & 5.44\\
\hline

\end{tabular}
\captionsetup{justification=centering,labelsep=period}
\caption{Optimized solar area scale factors by regions under different solar production capacity constraints}
\label{tab:optimization_results}
\end{table}

\begin{table}
\centering
\small
\begin{tabular}{ | L{0.7 cm} | L{2.2 cm}| L{2 cm}|L{2 cm}| } 
\hline
\textbf{Hrs} & \textbf{Consumption (In MW)} & \textbf{Production(In MW)} & \textbf{Production-Consumption} \\
\hline
1 & 49813  &  66449 &  16636\\
\hline
2 & 29990 & 53683 & 23693\\
\hline
3 & 43568 & 47961 & 4393\\
\hline
4 & 45010 & 45010 & 0\\
\hline
5 & 32106 & 32110 & 4\\
\hline
6 & 25547 & 38166 & 12618\\
\hline
7 & 36009 & 51664 & 15655\\
\hline
8 & 37574 & 47928 & 10354\\
\hline
9 & 57347 & 58913 & 1566\\
\hline
10 & 46638 & 46661 & 23\\
\hline
11 & 45316  &  49624 & 4309\\
\hline
12 & 38726 & 52790 & 14063\\
\hline
13 & 31021 & 57217 & 26196\\
\hline
14 & 35586 & 50006 & 14420\\
\hline
15 & 26953 & 44652 & 17699\\
\hline
16 & 48428 & 48431 & 3\\
\hline
17 & 50240 & 53595 & 3355\\
\hline
18 & 66904 & 66928 & 24\\
\hline
19 & 43897 & 77572 & 33675\\
\hline
20 & 49891 & 91494 & 41604\\
\hline
21 & 42928 & 93238 & 50309\\
\hline
22 & 46343 & 87249 & 40906\\
\hline
23 & 51142 & 90339 & 39197\\
\hline
24 & 43344 & 72324 & 28981\\
\hline
\end{tabular}
\captionsetup{justification=centering,labelsep=period}
\caption{Total consumption and production at each simulation hour (Experiment 1)}
\label{tab:Hourly production and consumption}
\end{table}

\begin{figure}[ht]
	\centering
	\captionsetup{justification=centering,labelsep=period}
	\includegraphics[width=.45\textwidth]{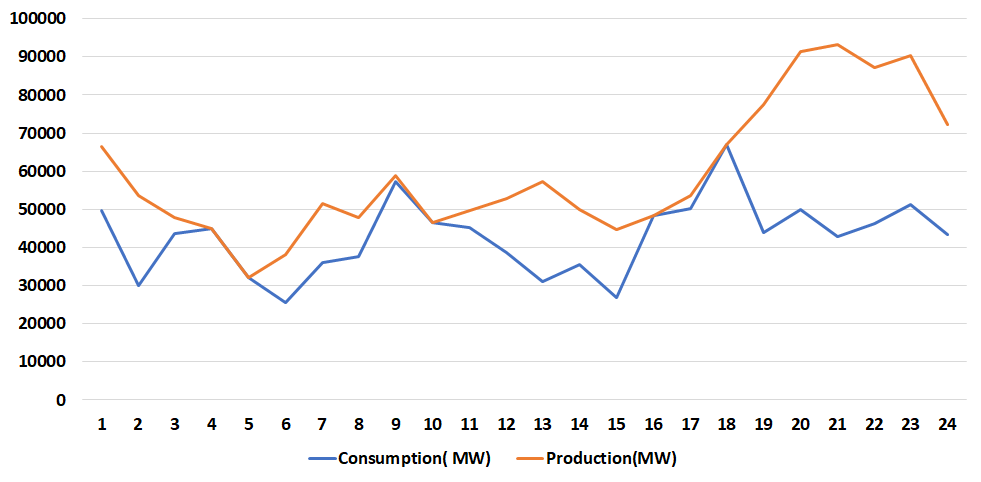}
	\caption{Total consumption and production (Experiment 1)}
	\label{fig:exp1}
\end{figure}

The output of the optimization was the scale factor. The scale factor was multiplied by the unitized area of a solar panel to obtain the solar panel area required at each location.  The optimization result of experiment 1 indicates that some of the locations in the model required zero solar area. This is due to the overlapping of solar energy production at different locations. In experiment 2, solar installation at each location for producing $40\%$ of that location's total consumption was required. Figure ~\ref{fig:exp1} shows the energy consumption and production pattern during the 24-hour period for experiment 1 (refer to table~\ref{tab:Hourly production and consumption} for data). The total energy produced in the network for this duration was always greater than the total energy consumption. During some parts of the day, production was significantly greater than consumption. 

\begin{figure}[ht]
	\centering
	\captionsetup{justification=centering,labelsep=period}
	\includegraphics[width=.45\textwidth]{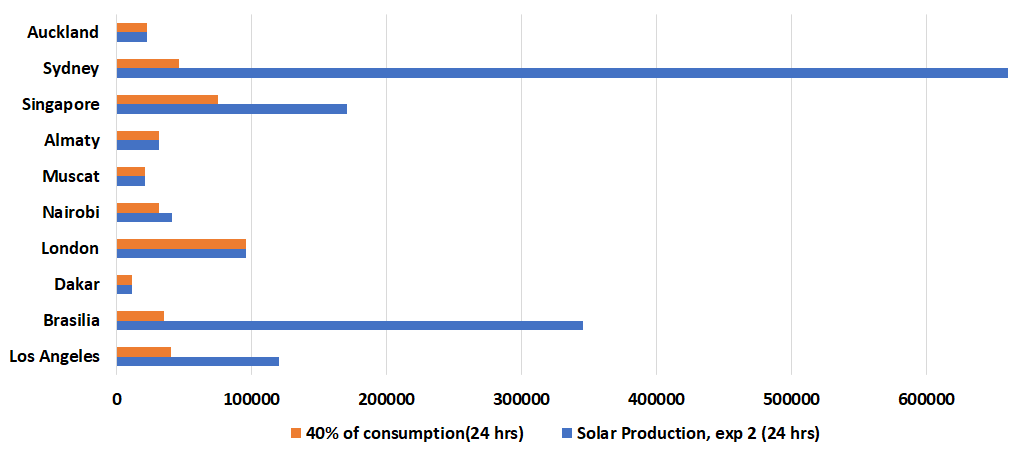}
	\caption{40\% of consumption and production (Experiment 2)}
	\label{fig:exp2}
\end{figure}

In experiment 2, we needed more solar area than in experiment 1 due to stricter constraints. Thus resulted in more unbalanced production. The production-consumption pattern for experiment 2 is shown in Figure~\ref{fig:exp2}.
Considering the 24 hour consumption for each location (Table~\ref{tab:Consumption_storage data}), the simulation data seem reasonable and closer to real data. Although city-level energy consumption data are not readily available, data for Los Angeles and Auckland were compared with available data and scaled by population. Los Angeles County's population is almost double that of Los Angeles City\cite{UScensus}. Energy consumption by Los Angeles County for 2017 was ~68800 GW\cite{CalEnergy}. Scaled down to the LA city population (division by 2) and to a one day average consumption (division by 365), this value becomes ~94 GW, which is close to  the 99 GW from the experimental data. We performed such comparisons to validate the approximation and accuracy of the models.
\begin{table}
\centering
\small
\begin{tabular}{ | L{1.8 cm} | L{2.8 cm}| L{2.6 cm}| } 
\hline
\textbf{Location} & \textbf{Consumption (24 hrs) (In GWh)} & \textbf{Storage(In GWh)} \\
\hline
Los Angeles & 99.9  &  62.20 \\
\hline
Bras\'{i}lia & 87.54 & 25.82 \\
\hline
Dakar & 28.10 & 14.83 \\
\hline
London & 240.27 & 169.11 \\
\hline
Nairobi & 77.75 & 37.72 \\
\hline
Muscat & 52.70 & 30.75\\
\hline
Almaty & 79.37 & 59.98 \\
\hline
Singapore & 187.24 & 70.21 \\
\hline
Sydney & 115.23 & 23.22 \\
\hline
Auckland & 56.17 & 11.77\\
\hline

\end{tabular}
\captionsetup{justification=centering,labelsep=period}
\caption{Total energy consumption and storage in absence of grid (Experiment 1)}
\label{tab:Consumption_storage data}
\end{table}
Energy storage capacity required at each location was also estimated in the absence of grid connectivity. Table~\ref{tab:Consumption_storage data} presents the minimum amount of energy storage required at each location. The estimation was performed by assuming that there is no power source other than solar, so total power required at night must be compensated by energy storage. The amount of storage required is higher than any existing installation in world\cite{Tesla}. 

Land availability is critical for solar panel installation. However, the average solar area per household (1898 $ft^2$) was approximately the average floor area of the household (1735 $ft^2$). This average assumed fixed-axis and 15\% single crystal silicon solar panel types, as well as high population density across all locations. Average solar area per household can be further improved with efficient solar panels and integration of sparser grid locations.
\section{Discussion}
\label{sec:discussion}
\setlength{\parskip}{0pt}
\setlength{\parsep}{0pt}
\setlength{\headsep}{0pt}
\setlength{\topskip}{0pt}
\setlength{\topmargin}{0pt}
\setlength{\topsep}{0pt}
\setlength{\partopsep}{0pt}
\captionsetup{justification=raggedright,singlelinecheck=false}

\subsection{GridLAB-D Design Studio}
The GridLAB-D Design Studio (DS)\cite{neema2019web} utilizes GridLAB-D simulator in the backend for modeling and simulation of the power grid. The DS provides a comprehensive web-based platform for analyzing distribution grids through simulations and evaluation of its resilience against network attacks.

\subsection{Optimization using Linear Programming}
Simulation output from the GridLAB-D DS was the total power produced by a unitized solar panel and power consumption by each house. This depended on the date of simulation, as weather data is based on a typical meteorological year. A single day simulation data was captured, then scaled to the number of houses to frame the optimization problem. The optimization problem minimized the area required for solar panels at different locations, while meeting energy needs at all locations. This was formulated as a linear programming (LP) problem. The LP problem was modeled in AMPL\cite{ample} - a mathematical programming language. The optimization model and simulation data were given to a solver called MINOS, which uses the \textit{Primal Simplex} algorithm for solving the LPs. The solver output was the solar panel area scale factor required at different locations. These panel area scale factors were multiplied by the unitized solar panel area to obtain actual required solar panel areas.

\subsection{Model of Integration and Power Exchange}
The chosen model of integration for a global grid and power exchange framed our optimization problems. In experiment 1, we assumed that any location could have any amount of solar area. Our goal was to find the required solar area, irrespective of panel distribution at each location. This assumption led to some locations with large area requirements, and other locations with zero solar panel area requirements (see Table~\ref{tab:optimization_results}). Experiments considering different models, backed by real power simulation data for long-term simulation, can give better insights into finalizing the model of integration for power exchange and solar panel installation. 

\subsection{Energy Storage Requirements in Absence of a Global Grid}
The Hornsdale Power Reserve\cite{Tesla}, which is connected to the Southern Australian grid, is the world's largest lithium-ion battery system. It contains hundreds of \textit{Tesla power packs}, each with 16 \textit{battery pods}. Each battery pod provides 129 MWh and houses thousands of small lithium-ion cells. This is significantly small compared to the amount of energy storage needed in the absence of a longitudinally-distributed global grid (refer Table~\ref{tab:Consumption_storage data}). Thus, grid integration provides a sustainable path to move toward renewable energy without significant additional storage.

\subsection{Critical Appraisal of the Experiment}
Our experiment was simplified to approximate a real world scenario and abstracted to optimize only for household consumption and solar production. Larger simulations must be performed at different abstraction levels for more reliable results and to better understand different facets of problem. Integration of different partners involved in the energy sector will give us a better perspective on the whole problem.   

\subsection{Selection of locations}
These experiments were use cases to validate how we can integrate the power distribution simulator with optimization tools. Our long term goal is to develop a generic toolchain through which users can select different locations and integrate them for their experiments. Interwoven networks with greater numbers of interconnected locations can give a better optimized results in terms of uniformly distributed solar panels in each location.

\subsection{Where is the transmission loss?} The experiment has limitations in terms of the transmission related losses, as we ignored such losses. This was done to simplify the problem and focus on integrating the distribution grid simulation with optimization tools. A richer experiment involving transmission grids requires a co-simulation platform such as C2WT-TE\cite{4756}, where each simulation forms a component of a larger simulation.

\subsection{Effect of weather} 
Weather affects the solar production as well as users' power consumption patterns. We simulated our experiments over a 24 hour time period. A more stable solution to finding required areas of solar panels at different locations can be found by optimizing over longer time periods and integrating more locations into the grid. We have weather data from approximately 30-40 years for most large cities in the world, and have good models for weather prediction as well. Optimization performed over a larger timescale considers broader variation in weather and consumption patterns, thereby yielding a more accurate result. Availability of GPUs and parallel computing can enable such computationally intensive experiments.

\subsection{Economic feasibility and Practicality}
It seems reasonable that migration to renewable energy sources cannot be achieved by relying on energy storage devices until there is a major breakthrough in storage capacity. Currently, the battery option is economically infeasible. Presently, the only feasible path towards clean energy is global grid integration. To achieve this integration, we need a tool to simulate scenarios and to see how the integrated grid will behave. Our experiments do not account for economic feasibility and are abstracted to calculate solar production and consumption patterns, with focus on user demand patterns at the household level and their dependence on weather. These global level power simulations need to be done at various levels of abstraction. A more robust and reliable grid can be designed by integrating various conventional and unconventional energy resources in unison where the majority of power is driven by renewable resources. 

\section{Conclusion and Future Work}
\setlength{\parskip}{0pt}
\setlength{\parsep}{0pt}
\setlength{\headsep}{0pt}
\setlength{\topskip}{0pt}
\setlength{\topmargin}{0pt}
\setlength{\topsep}{0pt}
\setlength{\partopsep}{0pt}
\captionsetup{justification=raggedright,singlelinecheck=false}
\label{sec:conclusionFutureWork}

\subsection{Conclusion}
 Experiments' result show that:
 \begin{enumerate}
     \item  By connecting ten large cosmopolitan cities, the energy needs of those cities can be met only by solar energy production.
     \item The area required for solar panels is moderate compared to the population density of these locations.
     \item The required battery capacity in absence of a global grid is enormous and economically infeasible. 
     \item We approximated models to imitate grid components, houses, appliances, and scheduling. Consumption patterns produced by the household level simulation matched real consumption data. 
 \end{enumerate}

\subsection{Future Work}
Our experiments are use cases to understand how we should proceed with the development of a toolchain. 
For a broader discussion of reliance on renewable energy,  we are working on developing a toolchain that will allow researchers to design various experiments at multiple fidelity levels.
The toolchain will be useful for:
\begin{enumerate}
    \item Modeling of different power grid networks and models of integration and power exchange.
    \item Modeling different house models with different appliances and configurable schedules, integrating different locations, and analyzing results. 
    \item Studying the local renewable integrated grid and its effects on the conventional power grid. 
    \item Enabling AI-based prediction models to predict future needs and control strategies for renewable energy production.

\end{enumerate}

\section{Acknowledgments}
This work at Vanderbilt University was supported in part by the PIRE program of NSF under award \#1743772 and by National Institute of Standards and Technology under award 70NANB18H269.

\bibliographystyle{IEEEtran}
\bibliography{references}

\end{document}